\documentclass[12pt]{article}
\pdfoutput=1

\usepackage[bulletsep]{collref}
\usepackage{amssymb,graphicx}
\usepackage[intlimits]{amsmath}
\usepackage{bbm}
\usepackage[small]{subfigure}

\usepackage{MnSymbol}

\makeatletter \@addtoreset{equation}{section} \makeatother

\makeatletter
\let\old@startsection=\@startsection
\let\oldl@section=\l@section
\renewcommand{\@startsection}[6]{\old@startsection{#1}{#2}{#3}{#4}{#5}{#6\mathversion{bold}}}
\renewcommand{\l@section}[2]{\oldl@section{\mathversion{bold}#1}{#2}}
\makeatother

\makeatletter
\let\old@makecaption=\@makecaption
\def\@makecaption{\small\old@makecaption}
\makeatother

\newcommand{\beq}{\stackrel{!}{=}}

\newcommand{\x}{{\tt x}}
\newcommand{\tp}{\top}
\newcommand{\K}{\mathcal{K}}

\begin{document}



\renewcommand{\thefootnote}{\fnsymbol{footnote}}
\setcounter{footnote}{0}

\begin{center}
{\Large\textbf{\mathversion{bold} String Integrability on the Coulomb Branch}
\par}

\vspace{0.8cm}

\textrm{Ron~Demjaha$^{1}$ and
Konstantin~Zarembo$^{2,3}$}
\vspace{4mm}

\textit{${}^1$Department of Physics and Astronomy, Uppsala University,
Box 516, SE-751 20 Uppsala, Sweden}\\
\textit{${}^2$Nordita, KTH Royal Institute of Technology and Stockholm University,
Hannes Alfv\'ens v\"ag 12, 106 91 Stockholm, Sweden}\\
\textit{${}^3$Niels Bohr Institute, Copenhagen University, Blegdamsvej 17, 2100 Copenhagen, Denmark}\\
\vspace{0.2cm}
\texttt{ron.demjaha.8644@student.uu.se, zarembo@nordita.org}

\vspace{3mm}


\par\vspace{1cm}

\textbf{Abstract} \vspace{3mm}

\begin{minipage}{13cm}
 The Coulomb branch of the $N=4$ super-Yang-Mills theory is described by a D3-brane in the bulk of $AdS_5\times S^5$. We show  that the boundary conditions on the D-brane preserve integrability of the  string sigma-model by  constructing a dynamical and spectral parameter dependent reflection matrix that encodes an infinite number of conserved charges.
\end{minipage}

\end{center}

\vspace{0.5cm}


\setcounter{page}{1}
\renewcommand{\thefootnote}{\arabic{footnote}}
\setcounter{footnote}{0}

\section{Introduction}

\begin{figure}[t]
 \centerline{\includegraphics[width=6cm]{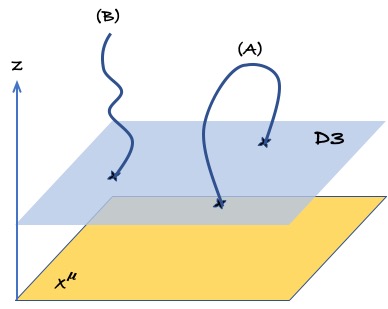}}
\caption{\label{StringPic}\small D3-brane dual of the Coulomb branch.}
\end{figure}

We demonstrate following \cite{Dekel:2011ja,Linardopoulos:2021rfq} that a lateral D3-brane in Anti-de-Sitter space (fig.~\ref{StringPic}) preserves integrability of the string sigma-model. The holographic dual of this setup is the Coulomb branch of the $\mathcal{N}=4$ super-Yang-Mills theory with the  $SU(N+1)$ symmetry broken to $SU(N)\times U(1)$.
This is perhaps the simplest holographic model of massive particles with a well-defined S-matrix and normal cuts and poles in the Green's functions.

Integrability on the Coulomb branch has been studied from various angles \cite{Alday:2009zm,Caron-Huot:2014gia,Loebbert:2020hxk,Caron-Huot:2021usw,Ivanovskiy:2024vel,Coronado:2025xwk}, including a remarkable conjecture about the non-per\-tur\-ba\-tive spectrum of bound states \cite{Caron-Huot:2014gia} and exact results for the one-point functions \cite{Coronado:2025xwk}. In all this work integrability of the Coulomb branch was taken for granted, as inherited from the conformal phase of the theory. While the results do suggest that integrability is preserved, {\it a priori} this is not guaranteed and requires justification. Reasons to question integrability of the Coulomb branch also exist.

The AdS/CFT integrability is more transparent when viewed from the string perspective. The equations of motion for the string on $AdS_5\times S^5$ admit Lax representation \cite{Bena:2003wd} and hence possess an infinite number of conservation laws which preclude particle production on the string worldsheet and open an avenue for powerful algebraic methods of Bethe Ansatz to study spectral data and correlation function.

The boun\-dary conditions for an open string attached to a D-brane generically break integrability, but may also preserve it in special cases, which for the string in $AdS_5\times S^5$ were classified in \cite{Dekel:2011ja}. Quite surprisingly, the Coulomb-branch D3-brane does not appear in this classification. Moreover, for strings ending on  a circular  brane in $AdS_3$, a close proxy of the Coulomb-branch brane in $AdS_5$,  integrability is actually broken \cite{Ishii:2023qqd}. The question therefore calls for further investigation.

The analysis of  integrability conditions in  \cite{Dekel:2011ja} is complete and exhaustive under certain assumptions. For once, the reflection off the D-brane is modeled in  \cite{Dekel:2011ja} by a constant matrix with numerical entries which is quite a natural assumption, but exceptions to it are also known. For example, integrability conditions for the Karch-Randall D5-brane admit more solutions if the reflection matrix is allowed to be dynamical  \cite{Linardopoulos:2021rfq}, dependent on the worldsheet fields. Short of a constant reflection matrix for the Coulomb branch, we are going to search for a dynamical one  in this  case  as well. In the next section we introduce the  formalism of double-row transfer-matrices \cite{Sklyanin:1987bi}, extended to accommodate dynamical reflection factors and specifically tailored for $AdS_5$  \cite{Linardopoulos:2021rfq} (a more detailed review can be found in~\cite{Linardopoulos:2025ypq}).  In sec.~\ref{D3-sec} we apply this formalism to the Coulomb-branch D3-brane in fig.~\ref{StringPic}.

\section{Integrable boundary conditions in $AdS_5$}

The classical bosonic string on $AdS_5$ in the conformal gauge
 is defined by the Lagrangian:
\begin{equation}\label{Poincare-lag}
 \mathcal{L}=\frac{\left(\partial_\alpha  x^\mu \right)^2+\left(\partial_\alpha  z\right)^2}{2z^2}\,.
\end{equation}
Adding  $S^5$ and fermions is trivial at the classical level and will be discussed later. Fixing the conformal gauge is not really important for our purposes, just streamlines the notations: the equations of motion, when expressed in differential forms,  look the same for any 2d metric.

The boundary conditions for the string ending on the D3-brane in fig.~\ref{StringPic} are Dirichlet for $z$ and Neumann for $x^\mu $:
\begin{equation}\label{DN-bc's}
 \dot{z}\beq 0,\qquad \acute{x}^\mu \beq 0,
\end{equation}
where the dot denotes derivative in $\tau $, the prime derivative in $\sigma $, and $\beq$ stands for equality at the string's endpoints. Integrability conditions are local, so instead of the string with both ends anchored to the brane (fig.~\ref{StringPic}A), we consider a semi-infinite string in fig.~\ref{StringPic}B. Again, this is just a matter of convenience and we later comment on the open string of a finite length.

\subsection{$AdS_5$ as a coset}

Geometrically, $AdS_5$ is a homogeneous space of the conformal group $SO(4,2)$. The standard coset construction \cite{Callan:1969sn}  then immediately guarantees integrability \cite{Eichenherr:1979ci}, taking into account that $AdS_5=SO(4,2)/SO(4,1)$ is a symmetric space  (see \cite{Arutyunov:2009,Zarembo:2017muf} for a review).  Using the isomorphism $\mathfrak{so}(4,2)\simeq \mathfrak{su}(2,2)$, the generators of $\mathfrak{so}(4,2)$ can be represented by the 5d Dirac matrices $\gamma _a$  and their commutators $\gamma _{ab}$, all defined  in the $(-++++)$ signature \cite{Arutyunov:2009}. The standard conformal generators are identified with
\begin{equation}
 D=\frac{\gamma _4}{2}\,,
 \qquad 
 P_\mu =\Pi _+\gamma _\mu ,
 \qquad 
 K_\mu =\Pi _-\gamma _\mu ,
 \qquad 
 L_{\mu \nu }=\gamma _{\mu \nu },
\end{equation}
where $\mu,\nu =0\ldots 3$ are the Lorentz indices and
\begin{equation}
 \Pi _\pm=\frac{1\pm\gamma _4}{2}
\end{equation}
are the usual chiral projectors (we denote $\gamma ^5$ by $\gamma ^4$ for consistency with the 5d index bookkeeping).
The denominator subalgebra is generated by $\mathfrak{h}=\left<\gamma ^{ab}\right>$, and together with its orthogonal complement $\mathfrak{f}=\left<\gamma ^a\right>$  defines a $\mathbbm{Z}_2$ decomposition of  the coset: $\mathfrak{so}(4,2)=\mathfrak{h}\oplus\mathfrak{f}$.

The coset parameterization by the standard Poincar\'e coordinates starts with the $SO(4,2)$ group element
\begin{equation}\label{cosrep}
 g=\,{\rm e}\,^{P_\mu x^\mu }z^D.
\end{equation}
The gauge-covariant (moving-frame) current defines the coset decomposition:
\begin{equation}
 J= g^{-1}dg\equiv A+K,\qquad A\in \mathfrak{h},~K\in\mathfrak{f}.
\end{equation}
In the explicit Poincar\'e parameterization,
\begin{equation}\label{AKA}
 A=\frac{1}{2z}\,dx^\mu \,\gamma _4\gamma _\mu ,
 \qquad 
 K=\frac{1}{2z}\left(dz\,\gamma _4+dx^\mu \,\gamma _\mu \right).
\end{equation}
It is then easy to check that the action of the coset sigma-model
\begin{equation}
 \mathcal{L}=\frac{1}{2}\,\mathop{\mathrm{tr}}K\wedge *K
\end{equation}
coincides with the string Lagrangian (\ref{Poincare-lag}). 

The equations of motion of  the coset can be compactly written as  
\begin{equation}
 F+K\wedge K=0,\qquad  DK=0,\qquad D*K=0,
\end{equation}
where $D$ is the covariant derivative: $DC=dC+A\wedge C +C\wedge A$, and $F$ is the curvature of the gauge connection: $F=dA+A\wedge A$.
We can also define the gauge-invariant (fixed-frame) current
\begin{equation}
 j=gKg^{-1}
\end{equation}
and re-write the equations of motion in a gauge-invariant form:
\begin{equation}
 dj-2j\wedge j=0,\qquad d*j=0.
\end{equation}
The fixed-frame current can be easily derived from (\ref{cosrep}) and  (\ref{AKA}):
\begin{equation}\label{j-current}
 j=\frac{1}{2z^2}\left[
 2(zdz+xdx)(D-xP)+(z^2+x^2)P dx+ K dx +L_{\mu \nu }x^\mu dx^\nu 
 \right],
\end{equation}
where $x^2=x^\mu x_\mu $,  $xdx\equiv x^\mu dx_\mu $ and so on.

The equations of motion can be further packaged into the flatness condition for the Lax connection:
\begin{equation}\label{a(x)}
 a(\x)=2\,\frac{j-\x*j}{1-\x^2}\,.
\end{equation}
The current so defined has zero curvature for any value of the spectral parameter $\x$:
\begin{equation}
 da+a\wedge a=0.
\end{equation}
The flatness condition is equivalent to the equations of motion and at the same time encodes an infinite number of conservation laws.

The generating function for the conserved charges is the monodromy matrix:
\begin{align}
    \mathcal{M}(\tau ;\mathrm{x})  = {\rm P}\,\mathrm{exp}\left( \int^{\infty}_0 \mathrm{d}s \ a_{\sigma}(\tau ,s;\mathrm{x}) \right)
\end{align}
The monodromy of a flat connection does not change under smooth deformations of the contour. As a consequence, the monodromy matrix is time-independent, provided the currents satisfy the equations of motion. This is literally true for an infinite worldsheet extending from $-\infty $ to $+\infty $. The trace monodromy for a closed string would also be automatically conserved. But an endpoint of the open string is pinned to the D-brane and cannot move, necessitating extra steps in the  construction of the conserved quantities.

\subsection{Double-row transfer matrix}

The generating function of conserved charges for an open string is defined by the folding trick \cite{Sklyanin:1987bi}:
\begin{align}\label{transferT}
    T(\mathrm{x}) = \mathcal{M}^\tp(-\mathrm{x})\mathbbm{U}(\mathrm{x})\mathcal{M}(\mathrm{x}).
\end{align}
The monodromy runs along the string back and forth, with the reflection matrix inserted in between. Transposition reverts path-ordering and hence orientation. The change of sign of the spectral parameter serves the same purpose\footnote{The spectral parameter behaves as a pseudoscalar because in (\ref{a(x)}) it multiplies $*j$, a pseudo-vector.}. We allow the reflection matrix to be a function of the spectral parameter  and also to depend on the dynamical variables $x^\mu (\tau ,0)$ and $z(\tau ,0)$\footnote{The $z$ coordinate is constant along the brane so the reflection matrix in effect only depends on $x^\mu $. Unless we want to compare boundary conditions for  D-branes at different values of $z$.}.

The time derivative of the monodromy matrix $\mathcal{M}$ picks a contribution from the string's endpoint:
\begin{align}
    \Dot{\mathcal{M}}(\mathrm{x}) = -a_{\tau}(0;\mathrm{x})\mathcal{M}(\mathrm{x}).
    \label{monodromy_derivative}
\end{align}
This is true for a semi-infinite string, the finite interval will be discussed later. The transfer matrix defined in (\ref{transferT}) will thus be time-independent provided the reflection matrix satisfies
\begin{equation}\label{icon}
 \dot{\mathbbm{U}}(\x)\beq a^\tp_\tau (-\x)\mathbbm{U}(\x)+\mathbbm{U}(\x)a_\tau (\x).
\end{equation}
If it is possible to find $\mathbbm{U}(\x)$ such that this equation  turns into identity, the boundary conditions will preserve integrability by construction.

The discussion so far was completely general, not specific to  the sigma-model on $AdS_5$. This latter case carries an extra algebraic structure.
Transposition, generically an outer automorphism of the symmetry algebra, for $\mathfrak{so}(4,2)$ is related to an inner automorphism combined with the $\mathbbm{Z}_2$ parity of the coset, namely there exists a matrix $\K$ \cite{Arutyunov:2009} such that 
\begin{equation}\label{transposition-Dirac-matrices}
 \gamma _a^\tp=\K^{-1}\gamma _a\K,\qquad \gamma ^\tp_{ab}=-\K^{-1}\gamma _{ab}\K, \qquad \K^\tp=-\K.
\end{equation}
In the chiral representation of the Dirac matrices $\K=\gamma _{13}$.  This suggests to define a twisted reflection matrix
\begin{equation}   \label{KU}
 \widehat{\mathbbm{U}}=\K\mathbbm{U}
\end{equation}
and the transposition bracket \cite{Linardopoulos:2021rfq}:
\begin{equation}
   \left\langle A,B\right\rangle_{\pm} \equiv \mathcal{K}A^\tp\mathcal{K}^{-1} B\pm BA.
    \label{transp_brackets}
\end{equation}

Taking into account the explicit form of the Lax connection (\ref{a(x)}), the integrability condition becomes
\begin{align}
    \Dot{\widehat{\mathbbm{U}}}(\mathrm{x}) \beq \frac{2}{\mathrm{x}^2-1}\left(\left\langle j_{\tau}, \widehat{\mathbbm{U}}(\mathrm{x})\right\rangle_+ + \mathrm{x}\left\langle j_{\sigma}, \widehat{\mathbbm{U}}(\mathrm{x})\right\rangle_- \right).
    \label{int_cond}
\end{align}
This is the equation we are going to study.

\section{Reflection matrix for D3-brane}
\label{D3-sec}

Our strategy to solve for the reflection matrix is based on the following observation. The integrability condition (\ref{int_cond}) is linear in derivatives, with time derivatives appearing in $\dot{\widehat{\mathbbm{U}}}$ and $j_\tau $, and spacial derivatives in  $j_\sigma $. The  last term in (\ref{int_cond}) is the only place where $\sigma $-derivatives can reside, and  should thus vanish on  its own:
\begin{equation}
 \left\langle j_{\sigma}, \widehat{\mathbbm{U}}(\mathrm{x})\right\rangle_-\beq 0.
\end{equation}
Fortunately the $\sigma $ component of the current   (\ref{j-current}) takes a very simple form as a consequence of the Dirichlet-Neumann boundary conditions (\ref{DN-bc's}):
\begin{equation}
 j_\sigma \beq \frac{z'}{z}\left(D-xP\right).
\end{equation}
The integrability constraint thus becomes an algebraic condition
\begin{equation}\label{br-condition}
 \left\langle D-xP, \widehat{\mathbbm{U}}(\mathrm{x})\right\rangle_- = 0.
\end{equation}
We are going to enforce this condition first and then adjust the remaining freedom to account for time dependence in (\ref{int_cond}).

To begin with, we can write a general ansatz consistent with the symmetries of the problem:
\begin{align}
    \widehat{\mathbbm{U}} = a\Pi_- + b\Pi_+ + c\Pi_-\gamma_{\mu}x^{\mu} + d\Pi_+\gamma_{\mu}x^{\mu}
    \label{U_ansatz}
\end{align}
where the coefficients $a$, $b$, $c$ and $d$ can depend on $x^2$ and $z^2$, the latter just being a constant.

Next we insert (\ref{U_ansatz}) into (\ref{br-condition}). The transposition brackets between various combination of Dirac matrices were tabulated in \cite{Linardopoulos:2021rfq} and can be found there, but  for completeness we list the key formulas in the Appendix~\ref{trBrackets}. Using these results we find that in order for (\ref{br-condition}) to hold the coefficients of (\ref{U_ansatz}) must satisfy
$d=c$ and $b=-c$, with no constraints on $a$. That is to say,
\begin{equation}
  \widehat{\mathbbm{U}} = a\Pi_- + c\left(\gamma _\mu x^\mu -\Pi _+\right).
\end{equation}
The integrability condition (\ref{int_cond}) then boils down to 
\begin{align}
    \Dot{\widehat{\mathbbm{U}}}(\mathrm{x}) \beq \frac{2}{\mathrm{x}^2-1}\,\left\langle j_{\tau}, \widehat{\mathbbm{U}}(\mathrm{x})\right\rangle_+ ,
    \label{int_cond-simplified}
\end{align}
with
\begin{equation}
 j_\tau =\frac{1}{2z^2}\left[
 2x\dot{x}\left(D-xP\right)+\left(z^2+x^2\right)\dot{x}P+\dot{x}K+L_{\mu \nu }x^\mu \dot{x}^\nu 
 \right].
\end{equation}
Using again the results in the Appendix~\ref{trBrackets} we find three consistency conditions:
\begin{align}
 \dot{c}&=0
\nonumber \\
 c&=\frac{a+c\left(x^2-z^2\right)}{z^2\left(\x^2-1\right)}
\nonumber \\
\dot{a}&=-2cx\dot{x},
\end{align}
which are solved by
\begin{equation}
 c=-1,\qquad a=x^2-\x^2 z^2.
\end{equation}

As a result, we find the following reflection matrix:
\begin{equation}
 \widehat{\mathbbm{U}}=\left(x^2-\x^2z^2\right)\Pi _--x^\mu \gamma _\mu +\Pi _+.
\end{equation}
Its insertion at the string's endpoint defines a double-row transfer matrix which is time-independent on the equations of motion and thus  generates an infinite number of conserved charges. This construction explicitly demonstrates integrability of the Coulomb-branch brane in $AdS_5$.

\subsection{$AdS_5\times S^5$}

Generalization to $AdS_5\times S^5$ is straightforward because the D3-brane is localized on $S^5$, in the conformal gauge the equations of motion for $AdS_5$ and $S^5$ are independent of one another, and the $O(6)$ sigma-model with the Dirichlet boundary conditions in well-known to be integrable \cite{Ghoshal:1994bc,Moriconi:1998gc}. The reflection matrix in that case is  just numeric  \cite{Aniceto:2017jor,Gombor:2018ppd}. Combining the $AdS_5$ and $S^5$ reflection matrices into a supermatrix
\begin{equation}
 \mathbbm{U}_{PSU(2,2|4)}=\begin{bmatrix}
 \mathbbm{U}_{AdS_5}   & 0 \\ 
  0  & \mathbbm{U}_{S^5}  \\ 
 \end{bmatrix},
\end{equation}
and using the Lax connection of the $PSU(2,2|4)/SO(4,1)\times SO(5)$ supercoset \cite{Bena:2003wd} we can define a double-row transfer matrix for the full-fledged Green-Schwarz sigma model on $AdS_5\times S^5$.

\subsection{Open string}

So far we have discussed a semi-infinite string with one end attached to the D3-brane. A more physically relevant case is a string stretched between two different D-branes or ending on the same D-brane. The double-row transfer matrix in this case is also constructed by the folding trick without a need of new ingredients. The integrability condition is local, so the addition of another endpoint does not affect the analysis we performed for the first endpoint. 

The monodromy matrix for the entire string takes the form 
\begin{align}
    T(\mathrm{x}) =\mathop{\mathrm{tr}} \mathcal{M}^\tp(-\mathrm{x)}\mathbbm{U}(0;\mathrm{x})\mathcal{M}(\mathrm{x})\mathbbm{U}^{-1}(\pi ;\mathrm{x})
\end{align}
where the spacial worldsheet coordinate varies in the range $\sigma \in [0,\pi]$, and we explicitly indicated the dependence of the reflection matrices on the endpoints where they are inserted. 

The time derivative  of the monodromy matrix $\mathcal{M}$ now gives rise to two boundary terms:
\begin{align}
    \Dot{\mathcal{M}}(\mathrm{x}) = \mathcal{M}(\mathrm{x})a_{\tau}(\pi;\mathrm{x}) - a_{\tau}(0;\mathrm{x})\mathcal{M}(\mathrm{x}),
\end{align}
and we find the following integrability condition for the $\sigma = \pi$ endpoint:
\begin{equation}
 \mathbbm{U}^{-1}(\x)\dot{\mathbbm{U}}(\x) \mathbbm{U}^{-1}(\x)\beq a_\tau (\x)\mathbbm{U}^{-1}(\x)+\mathbbm{U}^{-1}(\x)a_\tau ^\tp(-\x).
\end{equation}
But this is  equivalent to (\ref{icon}) and thus is solved by the same reflection matrix. Notice that cyclicity of the trace was imperative for arriving at this conclusion.

\section{Conclusions}

Albeit strong indications for integrability already exist we find it conceptually important to explicitly construct the hierarchy of conserved charges responsible for  integrability of the Coulomb branch. Just like for the Karch-Randall brane \cite{Linardopoulos:2021rfq}, the  reflection factor turns out to depend on the dynamical variables of the string and on the spectral parameter. What implications this  has for quantum integrability of the Coulomb branch is an interesting open question to which we have no immediate answer.

An  obvious application of string integrability is construction and classification of classical string solutions \cite{Kazakov:2004qf}. Our results open an avenue for applying integrability tools to spinning strings on the Coulomb branch, for example to the string solutions that describe hydrogen type bound states of massive W-bosons. These well-known solutions \cite{Kruczenski:2003be} that currently are only known numerically \cite{Kruczenski:2003be,Herzog:2008bp}.

\subsection*{Acknowledgements}
The work of K.Z. was supported by VR grant 2021-04578. 

\appendix

\section{Transposition brackets}\label{trBrackets}

The general form of the transposition brackets follows from the definition (\ref{transp_brackets}) and transposition properties of the Dirac  matrices  (\ref{transposition-Dirac-matrices}):
\begin{equation}
 \left\langle \gamma _a,\Gamma \right\rangle_\pm=[\gamma _a,\Gamma ]_\pm,\qquad \left\langle \gamma _{ab},\Gamma \right\rangle=-[\gamma _{ab},\Gamma ]_\mp,
\end{equation}
where $\Gamma $ stands for any $4\times 4$ matrix. Any transposition bracket can be derived from there, but for convenience we explicitly list particular cases  encountered in the main text:
\begin{align}
 &\left\langle \gamma _4,\Pi _\pm\right\rangle_-=0
\nonumber \\
&\left\langle \gamma _4,\Pi _\pm\gamma _\mu \right\rangle_-=\pm 2\Pi _\pm\gamma _\mu 
\nonumber \\
&\left\langle \Pi _+\gamma _\mu ,\Pi _+\right\rangle_-=-\gamma _4\gamma _\mu 
\nonumber \\
&\left\langle \Pi _+\gamma _\mu ,\Pi _-\right\rangle_-=0
\nonumber \\
&\left\langle \Pi _+\gamma _\mu ,\Pi _+\gamma _\nu \right\rangle_-
=\Pi _-\gamma _\mu \gamma _\nu 
\nonumber \\
&\left\langle \Pi _+\gamma _\mu ,\Pi _-\gamma _\nu \right\rangle_-
=-\Pi _-\gamma _\nu \gamma _\mu 
\nonumber \\
&\left\langle \gamma _4,\Pi _\pm\right\rangle_+=\pm 2\Pi _\pm
\nonumber \\
&\left\langle \gamma _4,\gamma _\mu \right\rangle_+=0
\nonumber \\
&\left\langle \Pi _\pm\gamma _\mu ,\Pi _\pm\right\rangle_+=\gamma _\mu
\nonumber \\
& \left\langle \Pi _\pm\gamma _\mu ,\Pi _\mp\right\rangle_+=0
\nonumber \\
&\left\langle \Pi _\pm\gamma _\mu ,\gamma _\nu \right\rangle_+=2\eta _{\mu \nu }\Pi _\mp
\nonumber \\
&\left\langle \gamma _{\mu \nu },\Pi _\pm\right\rangle_+=0
\nonumber \\
&\left\langle \gamma _{\mu \nu },\gamma _\lambda \right\rangle_+=2\eta _{\mu \lambda }\gamma _\nu -2\eta _{\nu \lambda }\gamma _\mu .
\end{align}

\bibliographystyle{nb}

\end{document}